\documentclass[sigconf, authorversion]{acmart}

\usepackage{balance}       %
\usepackage{graphics}      %
\usepackage{hyperref}
\usepackage{color}
\usepackage{booktabs}
\usepackage{textcomp}
\usepackage{lipsum}%
\usepackage{makecell}
\usepackage{multicol}
\usepackage{multirow}
\usepackage{array}
\usepackage{xparse}
\usepackage{verbatimbox}
\usepackage{enumitem}
\usepackage{amsmath}
\usepackage{stfloats}
\usepackage{amsthm}
\usepackage{listings}
\usepackage{caption} 
\usepackage{xspace}
\usepackage[linesnumbered]{algorithm2e}
\usepackage{algpseudocode}
\usepackage{tabularx}
\usepackage{arydshln}
\usepackage[bottom]{footmisc}
\usepackage{stackengine}
\usepackage{placeins}
\usepackage{graphicx}
\usepackage{subcaption} %
\usepackage{pifont}
\usepackage{colortbl}
\usepackage{xfrac}
\usepackage{makecell}

\newcommand*{\eg}{\textit{e.g.},\xspace}
\newcommand*{\ie}{\textit{i.e.},\xspace}
\newcommand*{\vs}{\textit{vs.}\xspace}

\newcolumntype{L}[1]{>{\raggedright\let\newline\\\arraybackslash\hspace{0pt}}m{#1}}
\newcolumntype{C}[1]{>{\centering\let\newline\\\arraybackslash\hspace{0pt}}m{#1}}
\newcolumntype{R}[1]{>{\raggedleft\let\newline\\\arraybackslash\hspace{0pt}}m{#1}}

\makeatletter
\def\thickhline{%
  \noalign{\ifnum0=`}\fi\hrule \@height \thickarrayrulewidth \futurelet
   \reserved@a\@xthickhline}
\def\@xthickhline{\ifx\reserved@a\thickhline
               \vskip\doublerulesep
               \vskip-\thickarrayrulewidth
             \fi
      \ifnum0=`{\fi}}
\makeatother

\makeatletter
\def\thickhlinespace{%
  \addlinespace[1ex]
  \noalign{\ifnum0=`}\fi\hrule \@height \thickarrayrulewidth \futurelet
   \reserved@a\@xthickhline
   \addlinespace[1ex]
   }
\def\@xthickhlinespace{\ifx\reserved@a\thickhline
               \vskip\doublerulesep
               \vskip-\thickarrayrulewidth
             \fi
      \ifnum0=`{\fi}}
\makeatother

\newlength{\thickarrayrulewidth}
\newlength\Origarrayrulewidth

\algnewcommand{\IfThenElse}[3]{%
  \State \algorithmicif\ #1\ \algorithmicthen\ #2\ \algorithmicelse\ #3}

\NewDocumentCommand{\statP}{O{=} m}{%
  \textit{p} #1 {#2}%
}

\definecolor{revisionColor}{HTML}{025DF4}

\newcommand{\cding}[2]{%
  \begingroup
  \definecolor{tmpCircColor}{HTML}{#1}%
  \textcolor{tmpCircColor}{\ding{\numexpr181 + #2\relax}}%
  \endgroup
}

\definecolor{teaserGreen}{HTML}{004D40}
\definecolor{teaserYellow}{HTML}{825500}
\definecolor{teaserBlue}{HTML}{1558B0}

\AtBeginDocument{%
  }

\copyrightyear{2026}
\acmYear{2026}
\setcopyright{cc}
\setcctype{by-nc-nd}
\acmConference[CHI EA '26]{Extended Abstracts of the 2026 CHI Conference on Human Factors in Computing Systems}{April 13--17, 2026}{Barcelona, Spain}
\acmBooktitle{Extended Abstracts of the 2026 CHI Conference on Human Factors in Computing Systems (CHI EA '26), April 13--17, 2026, Barcelona, Spain}
\acmDOI{10.1145/3772363.3798356}
\acmISBN{979-8-4007-2281-3/2026/04}

\begin{document}

\title{Graphing Inline: Understanding Word-scale Graphics Use in Scientific Papers}

\author{Siyu Lu}
\authornote{Both authors contributed equally to this research.}
\affiliation{%
  \institution{Tongji University}
  \city{Shanghai}
  \country{China}}
\email{2250898@tongji.edu.cn}

\author{Yanhan Liu}
\authornotemark[1]
\affiliation{%
  \institution{Tongji University}
  \city{Shanghai}
  \country{China}}
\email{2351591@tongji.edu.cn}

\author{Shiyu Xu}
\affiliation{%
  \institution{Independent Researcher}
  \city{Seattle}
  \state{WA}
  \country{USA}}
\email{shiyuxu99@gmail.com}

\author{Ruishi Zou}
\affiliation{%
  \institution{Tongji University}
  \city{Shanghai}
  \country{China}}
\email{zouruishi@tongji.edu.cn}

\author{Chen Ye}
\authornote{Chen Ye is the corresponding author.}
\affiliation{%
  \institution{Tongji University}
  \city{Shanghai}
  \country{China}}
\email{yechen@tongji.edu.cn}

\begin{abstract}
Graphics (e.g., figures and charts) are ubiquitous in scientific papers, yet separating graphics from text increases cognitive load in understanding text-graphic connections. Research has found that word-scale graphics, or visual embellishments at typographic size, can augment original text, making it more expressive and easier to understand. However, whether, if so, how scientific papers adopt word-scale graphics for scholarly communication remains unclear. To address this gap, we conducted a corpus study reviewing 909 word-scale graphics extracted from 126,797 scientific papers. Through analysis, we propose a framework that characterizes where (positioning), why (communicative function), and how (visual representation) authors apply word-scale graphics in scientific papers. Our findings reveal that word-scale graphics are rarely used, that icons dominate visual representation, and that visual representation connects with communicative function (e.g., using quantitative graphs for data annotation). We further discuss opportunities to enhance scholarly communication with word-scale graphics through technical and administrative innovations.
\end{abstract}



\begin{teaserfigure}
    \centering  
    \includegraphics[width=.95\linewidth]{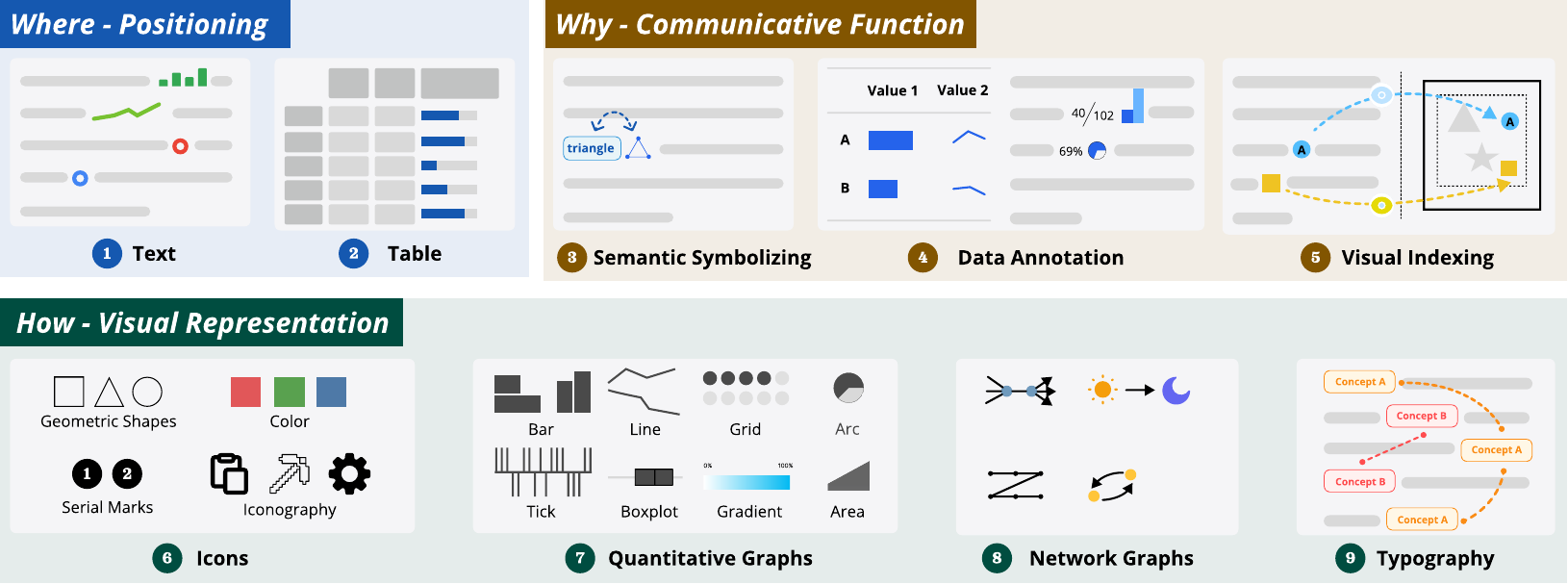} 
    \caption{We characterize the design space of word-scale graphics in scientific papers through a \textit{where-why-how} framework: \textit{\textbf{\textcolor{teaserBlue}{where}}} (Positioning: text or table), \textit{\textbf{\textcolor{teaserYellow}{why}}} (Communicative Function: visual indexing, semantic symbolizing, or data annotation), and \textit{\textbf{\textcolor{teaserGreen}{how}}} (Visual representation: icons, quantitative graphs, network graphs, and typography).}
    \label{fig:Taxonomy}
    \Description{}
\end{teaserfigure}

\maketitle

\section{Introduction}

Graphics such as figures and charts are central to scholarly communication, supporting the dissemination of knowledge and fostering understanding of scientific content~\cite{jessop2008Digital,huang2018Effect,oska2020Picture, guo2020You}. Given the affordance of graphics, authors would draft standalone graphics, typically positioned separately from the main text, to present data or advance arguments. Despite its benefits, connecting text to standalone graphics is difficult--readers must split their attention between the text and graphics to make sense of what the author wants to convey. This division of attention (\ie \emph{split-attention effect})~\cite{ayres2005SplitAttention} adds extraneous cognitive load on readers, hindering their comprehension of scientific papers~\cite{sweller1998Cognitive}. 

One method to reduce \emph{the split-attention effect} in reading is \emph{sparklines}~\cite{tufte2006beautiful}, defined as ``small, high-resolution graphics usually embedded in a full context of words, numbers, images.'' More recently, \citet{goffin2014Exploring,goffin2017Exploratory} generalized this idea to a family of \textbf{\textit{word-scale graphics}}--compact, miniature visuals that 1) span from letter height up to sentence/paragraph size, 2) support a variety of visual encodings, and 3) allow flexible placements inline and beyond. Following Goffin et al.'s terminology, we use \emph{word-scale graphics} to encompass the broad set of typographically sized visuals.

Prior research on word-scale graphics has mainly focused on its design and application, such as its design space ~\cite{goffin2014Exploring,brandes2013Gestaltlines,goffin2020Interaction,goffin2015Design,goffin2017Exploratory}, its impact to reading behavior~\cite{goffin2015Exploring}, and its novel applications~\cite{beck2016Visual,latif2019VIS,guo2020TopicBased,hoffswell2018Augmenting,blascheck2016Visual,lin2023Quest,perin2013SoccerStories,zou2025GistVis,lee2010SparkClouds}. Despite these advances, few have explored how word-scale graphics are applied ``in-the-wild''~\cite{goffin2017Exploratory,goffin2020Interaction}, especially in the context of scholarly communication~\cite{beck2017WordSized}. Perhaps the most closely related study is from \citet{beck2017WordSized}, who conducted a keyword-based literature search to assemble a corpus of scientific papers that included data-driven word-scale graphics (\ie representing numerical data). %
However, few have characterized how researchers use \textit{both data- and non-data-driven} (\ie represent text semantics) word-scale graphics across a real-world corpus of \textit{scientific papers}~\cite{goffin2017Exploratory,goffin2014Exploring}. 
Motivated by this gap, we propose our research question: \textbf{\textit{How do researchers apply word-scale graphics in scientific papers?}}

To address this research question, we assembled an in-the-wild corpus from 126,797 scientific papers\footnote{The majority portion of the papers comes from the year 2024.}. Through automatic extraction and manual annotation, we filtered out a dataset containing 909 distinct use cases of word-scale graphics,  aggregating all similar occurrences within each paper. By inductively analyzing this corpus, we derived a \emph{where-why-how} framework (\S\ref{sec:results}) that describes the positioning (\textit{where}), communicative function (\textit{why}), and visual representation (\textit{how}) word-scale graphics are applied in scientific papers. We also explored how the where-why-how dimensions associate with one another. Lastly, we connected our findings to prior work and translated them into design opportunities for authoring tools and publication workflows (\S\ref{sec:discussion}). 

In summary, our main contributions are:
\begin{itemize}
    \item We perform a corpus study on an in-the-wild corpus including 909 word-scale graphics use cases from scientific papers and derive a \textit{where-why-how} framework to characterize word-scale graphics use in scientific papers.
    \item Combining the corpus study and the framework, we discuss opportunities to enhance scholarly communication through word-scale graphics by leveraging technical and administrative innovations.
\end{itemize}

\section{Method}
\label{sec:corpus Collection}

\begin{figure*}[tbp]
    \centering  
    \includegraphics[width=0.95\linewidth]{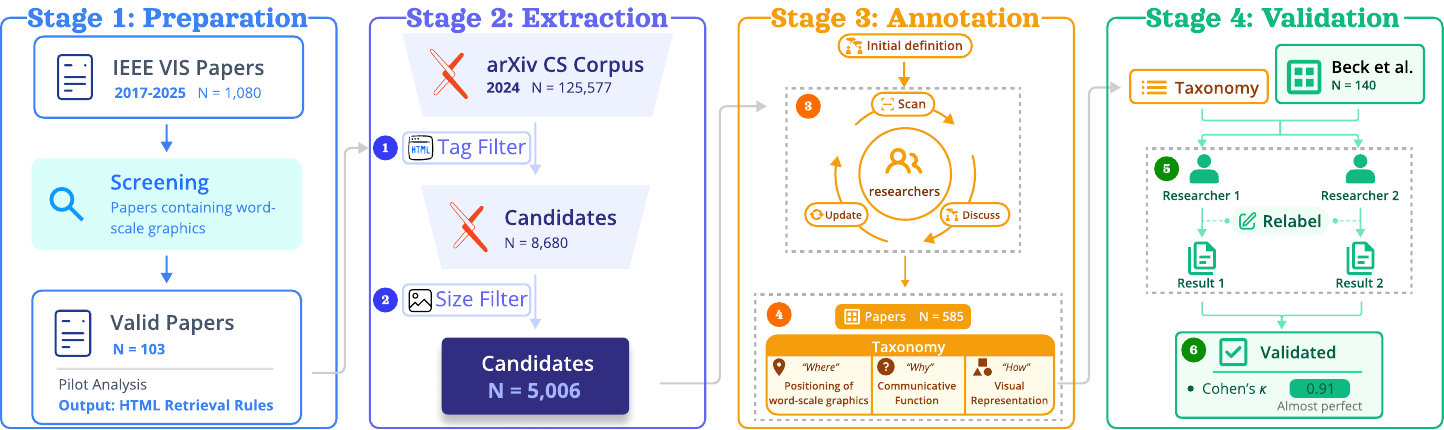} 
    \caption{The four-stage study pipeline. We 1) prepared retrieval rules based on 103 IEEE VIS papers; 2) extracted 5,006 candidates from 125,577 arXiv papers; 3) annotated 585 valid papers to derive the where (positioning), why (communicative function), and how (visual representation) framework; and 4) validated the framework on 140 papers from~\citet{beck2017WordSized}.}
    \label{fig:pipeline}
    \Description{}
\end{figure*}

We analyze word-scale graphics in scientific papers through four stages (Fig.~\ref{fig:pipeline}): Preparation (\textbf{S1}), Automatic Candidate Extraction (\textbf{S2}), Coding and Analysis (\textbf{S3}), and Validation (\textbf{S4}).

\paragraph{\textbf{S1: Preparation}.}
To obtain a heuristic for where and how word-scale graphics appear in scientific papers, we first collected a small corpus of papers published at the IEEE Visualization conference (IEEE VIS). We motivate this decision by assuming that visualization researchers are more likely to adopt word-scale graphics. Specifically, one author manually browsed every paper (\textit{N}=1,080) published in IEEE VIS between 2017-2025 in the IEEE Xplore digital library and identified 103 papers (\ie the \texttt{VIS} corpus, Fig.~\ref{fig:pipeline}.~\textbf{S1}). Besides gaining an initial understanding of the prevalence of word-scale graphics in scientific papers, we found that word-scale graphics are typically rendered via graphic-related tags (\eg \texttt{img}) nested within inline text containers (\eg \texttt{span}) that adhere to a height threshold (\eg 70px).

\paragraph{\textbf{S2: Automatic Candidate Extraction}.}
Having gained an understanding of the prevalence of word-scale graphics and how to capture them, we expanded our search scope to all arXiv Computer Science papers (category \texttt{cs}, \ie the \texttt{arXiv} corpus). We limit our scope to 2024 to ensure we have a full year's collection (the study was conducted in 2025), to keep the corpus as recent as possible, and to make it feasible to analyze in our study. We used an automated script that first crawled the arXiv HTML renders of all arXiv \texttt{cs} papers in 2024 (\textit{N}=125,577). Then, through two filters derived from the patterns observed in \textbf{S1} (\ie a tag filter (\texttt{span} tag) and a size filter enforcing (height < 70px)), Fig.~\ref{fig:pipeline}.~\textbf{S2} \cding{363AF5}{1}\&\cding{363AF5}{2}), we found 5,006 papers that likely included a word-scale graphic. During this automated extraction, we observed that graphics embedded in both text and tables share identical HTML structures.

\paragraph{\textbf{S3: Coding and Analysis}.}
\label{sec:Manual Annotation}

We started our corpus study with the \texttt{VIS} corpus to generate an initial set of codes to inform subsequent analysis. Three coders independently performed inductive coding to identify and characterize word-scale graphics in the corpus. Then, we synthesized individual codes through team discussions and formed an initial codebook. Subsequently, we scaled the coding process by applying the initial codebook to the 5,006 arXiv papers (Fig.~\ref{fig:pipeline}.~\textbf{S3} \cding{FC6D00}{3}). Two of the authors partitioned the candidates into two distinct sets and annotated them independently. During the process, we regularly discussed ambiguous cases and updated the definitions of each code. This process converged to a set of codes with two themes: 1) codes describing the \textit{positioning} and 2) codes describing \textit{communicative function} (Fig.~\ref{fig:pipeline}.~\textbf{S3} \cding{FC6D00}{4}) of word-scale graphics. In the end, we identified 585 papers containing word-scale graphics from the \texttt{arXiv} corpus. 

After finalizing the \texttt{arXiv} corpus, we combined \texttt{VIS} and \texttt{arXiv} corpora to further characterize \emph{how} word-scale graphics are visually represented--a dimension we did not focus on during the inductive coding. Drawing upon prior works~\cite{lohse1994Classification,goffin2017Exploratory} to propose an initial categorization, we inductively refined the dimensions and derived the specific sub-categories (Fig.~\ref{fig:Taxonomy}) to characterize our corpus.

\paragraph{\textbf{S4: Validation}.}
Two coders validated the framework by deductively coding 140 papers from the corpus compiled by \citet{beck2017WordSized} (\ie \texttt{Val} corpus, Fig.~\ref{fig:pipeline}.~\textbf{S4} \cding{0B8E04}{5}). We calculated the inter-rater reliability and derived a Cohen's $\kappa$ of $0.91$  (Fig.~\ref{fig:pipeline}.~\textbf{S4} \cding{0B8E04}{6})--a promising result~\cite{landis1977measurement} that shows the reliability of the framework.

\section{Results}
\label{sec:results}

The process in \S\ref {sec:corpus Collection} yielded codes for individual papers and a framework that characterizes word-scale graphics use in scientific papers. In this section, we first introduce our \emph{where-why-how} framework (\S\ref{sec:Overview}) and then explore the pairwise association between these dimensions (\S\ref{sec:Analysis}). We report our final results integrating our three data sources (\ie the \texttt{VIS}, \texttt{arXiv}, and \texttt{Val} corpora), including 718 unique papers (from 828 raw entries) and 909 distinct use cases. Note that only about $0.6\%$ of the 126,797 collected papers contain word-scale graphics, indicating that \textit{researchers rarely use word-scale graphics in their publications}. Such observation aligns with the finding of \citet{beck2017WordSized}, who reported scarce adoption of data-driven word-scale graphics in documents. We discuss the implications of this scarcity in \S\ref{sec:discussion}.

\subsection{Overview of the Framework}
\label{sec:Overview}

Our \textit{where-why-how} framework describes the application of word-scale graphics in scientific papers (Fig.~\ref{fig:Taxonomy}). In the following, we contextualize each subcategory with its prevalence in our corpus and papers that employed the corresponding techniques.

\subsubsection{\textcolor[HTML]{1558B0}{Where: Positioning of Word-scale Graphics}} We outline two types of word-scale graphics positioning:

\begin{enumerate}
  \item \emph{Text}~\protect\raisebox{-.2em}{\includegraphics[height=.9em]{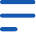}}~(\textit{N}=591, \protect\raisebox{-.2em}{\includegraphics[height=.9em]{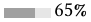}}): Graphics that appear embedded in text descriptions of scientific papers (\eg Fig.~\ref{fig:Taxonomy} \cding{1558B0}{1}, see more examples in \cite{cho2024Cocoon,hrynenko2025Identifyinga,bramas2024StandUp}). 
  
  \item \emph{Table}~\protect\raisebox{-.2em}{\includegraphics[height=.9em]{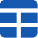}}~(\textit{N}=318, \protect\raisebox{-.2em}{\includegraphics[height=.9em]{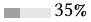}}): Graphics that appear embedded in tables of scientific papers (\eg Fig.~\ref{fig:Taxonomy} \cding{1558B0}{2}, see more examples in \cite{goffin2017Exploratory,wang2024KnowledgeSG,zahid2024Testing}).
\end{enumerate}

\subsubsection{\textcolor[HTML]{825500}{Why: Communicative Function}} We outline three types of communicative functions:

\begin{enumerate}[resume]
  \item \emph{Semantic Symbolizing}~\protect\raisebox{-.2em}{\includegraphics[height=.9em]{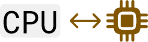}}~(\textit{N}=346, \protect\raisebox{-.2em}{\includegraphics[height=.9em]{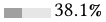}}): Graphics that serve as visual aids to symbolize specific concepts, thereby reinforcing the text description to enhance reader comprehension (Fig.~\ref{fig:Taxonomy} \cding{825500}{3}, see more examples in~\cite{chen2024RLGPT,beck2025PUREsuggest,rupp2025Level}).

  \item \emph{Data Annotation}~\protect\raisebox{-.2em}{\includegraphics[height=.9em]{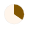}}~(\textit{N}=153, \protect\raisebox{-.2em}{\includegraphics[height=.9em]{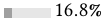}}): Graphics designed to allow direct extraction of data that directly encode quantitative data associated with texts (\eg  Fig.~\ref{fig:Taxonomy} \cding{825500}{4}, see more examples in~\cite{tanzil2024ChatGPT,das2024Why,ganesan2025Explaining}).

  \item \emph{Visual Indexing}~\protect\raisebox{-.2em}{\includegraphics[height=.9em]{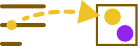}}~(\textit{N}=410, \protect\raisebox{-.2em}{\includegraphics[height=.9em]{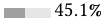}}): Graphics that serve the purpose of establishing a visual correspondence with a specific visual or textual entity mentioned in the text descriptions (\eg  Fig.~\ref{fig:Taxonomy} \cding{825500}{5}, see more examples in~\cite{zou2025GistVis,li2019Galex,wei2024SoK}).

\end{enumerate}

\subsubsection{\textcolor[HTML]{004D40}{How: Visual Representation}} We outline four broad categories for visual representations:

\begin{enumerate}[resume]
  \item \emph{Icons}~\protect\raisebox{-.2em}{\includegraphics[height=.9em]{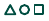}}~(\textit{N}=723, \protect\raisebox{-.2em}{\includegraphics[height=.9em]{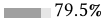}}): Graphics that typically adopt stylized visual forms of objects designed to evoke semantic associations (\eg Fig.~\ref{fig:Taxonomy} \cding{004D40}{6}, see more examples in \cite{zhao2024Optimizing,karamolegkou2025Evaluating,banse2024Owl2proto}). Subtypes of Icons may include geometric shapes, color boxes, serial marks, and iconography.

  \item \emph{Quantitative Graphs}~\protect\raisebox{-.2em}{\includegraphics[height=.9em]{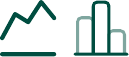}}~(\textit{N}=144, \protect\raisebox{-.2em}{\includegraphics[height=.9em]{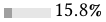}}): Graphics in which numeric data are encoded through certain visual channels (\eg length, angle), usually containing an implicit coordinate system (\eg Fig.~\ref{fig:Taxonomy} \cding{004D40}{7}, see more examples in \cite{hoffswell2018Augmenting,yu2024AntLM,alavi2024Game}). Examples include using visual primatives such as bar, line, grid, arc to represent data.

  \item \emph{Network Graphs}~\protect\raisebox{-.2em}{\includegraphics[height=.9em]{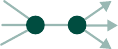}}~(\textit{N}=13, \protect\raisebox{-.2em}{\includegraphics[height=.9em]{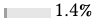}}): Graphics that represent relational structures such as node-link diagrams to enhance the perception of relationships (\eg Fig.~\ref{fig:Taxonomy} \cding{004D40}{8}, see more examples in \cite{qin2025Federated,jin2024Homomorphism,kneisel2024ToolAssisted}).

  \item \emph{Typography}~\protect\raisebox{-.2em}{\includegraphics[height=.9em]{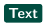}}~(\textit{N}=29,
  \protect\raisebox{-.2em}{\includegraphics[height=.9em]{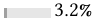}}): Graphics that encode texts with certain visual encodings (\eg color) to create consistent visual linkages within the document scope~\cite{brath2020visualizing} (\eg Fig.~\ref{fig:Taxonomy} \cding{004D40}{9}, see more examples in \cite{aniakor2024Survey,delemazure2024Comparing,zaman2025ChannelExplorer}).
  Specifically, we excluded simple text formatting (\eg bolding, highlighting)~\cite{strobelt2016Guidelines} as word-scale graphics because they might not be designed to create visual linkages across the document. Additionally, we categorized serial marks as icon type rather than typography because they are auxiliary to the main text, making them semantically more similar to icons than to text.

\end{enumerate}

Overall, word-scale graphics appear more often in text than tables (\protect\raisebox{-.2em}{\includegraphics[height=.9em]{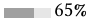}} \vs~\protect\raisebox{-.2em}{\includegraphics[height=.9em]{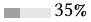}}), are mainly used for visual indexing and semantic symbolizing (\protect\raisebox{-.2em}{\includegraphics[height=.9em]{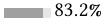}}), and most commonly occur in the form of icons (\protect\raisebox{-.2em}{\includegraphics[height=.9em]{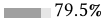}}).

\subsection{Correlation Analysis}  
\label{sec:Analysis}

To understand how the \textit{where}, \textit{why}, and \textit{how} dimensions interact with one another, we conducted Pearson's Chi-Square tests. In the following, we report $\chi^2$ values, \textit{p}-value, and Cramer's $V$ effect size.

\subsubsection{Association} We first report pairwise associations across dimensions. Our analyses reveal consistent patterns linking positioning, communicative function, and visual representation: all pairwise relationships are statistically significant with moderate-to-substantial effect sizes, suggesting that these dimensions co-occur in structured ways.

\emph{Association between ``where'' and ``why''.}
We observe a substantial association between the positioning of word-scale graphics and communicative function ($\chi^2(2)=340.56$, $p < 0.001$, $V=0.612$), suggesting that the embedding context is closely linked to the graphics' intended utility. Specifically, graphics embedded in text are primarily used for visual indexing (\textit{N}=391, \protect\raisebox{-.2em}{\includegraphics[height=.9em]{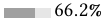}} within text), serving as visual anchors to support navigation and cross-referencing. In contrast, usage within tables tends to link with analytical purposes. semantic symbolizing emerges as the most frequent function in the table context (\textit{N}=177, \protect\raisebox{-.2em}{\includegraphics[height=.9em]{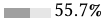}} within tables), followed by data annotation (\textit{N}=122, \protect\raisebox{-.2em}{\includegraphics[height=.9em]{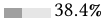}} within tables). One possible explanation of this pattern could be that the inherent structure of tables made them more favorable in presenting concepts and quantitative insights than the unstructured text.

\emph{Association between ``why'' and ``how''.}
Our analysis indicates a substantial association between the intended communicative function and the choice of visual representation ($\chi^2(6)=498.53$, $p < 0.001$, $V=0.524$). Based on such evidence, we hypothesize that communicative goals influence visual representation design. For instance, the goal of data annotation frequently links to the use of quantitative graphs (\textit{N}=113, \protect\raisebox{-.2em}{\includegraphics[height=.9em]{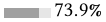}} within data annotation) to encode numerical values. Meanwhile, icons dominate both the visual indexing and semantic symbolizing goals. However, the alternative encodings for the above two communicative goals differ: typography ranked second for visual indexing, whereas authors use diverse encoding strategies for semantic symbolizing.

\emph{Association between ``where'' and ``how''.}
Our analysis reveals a moderate association between the positioning of word-scale graphics and visual representation strategies ($\chi^2(3)=82.38$, $p < 0.001$, $V=0.301$). We found that icons dominate visual representation in both table and text. Authors are also more likely to put data-driven word-scale graphics in tables--in \protect\raisebox{-.2em}{\includegraphics[height=.9em]{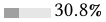}} (\textit{N}=98) of all use cases, quantitative graphs appear within tables, whereas only \protect\raisebox{-.2em}{\includegraphics[height=.9em]{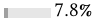}} (\textit{N}=46) appear within texts.

\subsubsection{Dominant usage patterns.}

By examining the joint distribution of the three dimensions, we identified the five most prevalent usage combinations, which together account for the majority of word-scale graphics in our corpus. Icons dominate the visual representation of word-scale graphics in scientific papers, appearing in 4 of the 5 most prevalent usage combinations. For example, ``text + visual indexing + icon'' accounts for \protect\raisebox{-.2em}{\includegraphics[height=.9em]{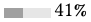}}, ``table + semantic symbolizing + icon'' accounts for \protect\raisebox{-.2em}{\includegraphics[height=.9em]{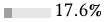}}, ``text + semantic symbolizing + icon'' accounts for \protect\raisebox{-.2em}{\includegraphics[height=.9em]{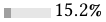}}, ``table + data annotation + icon'' accounts for \protect\raisebox{-.2em}{\includegraphics[height=.9em]{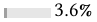}}. Additionally, a prominent pattern without icons is ``table + data annotation + quantitative graphs'', which accounts for \protect\raisebox{-.2em}{\includegraphics[height=.9em]{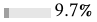}}. Overall, while icons demonstrate broad versatility across diverse functions and contexts, quantitative graphs have a strong presence in tables--likely because the structures of tables benefit data presentation.

\section{Discussion}
\label{sec:discussion}

Building on the \textit{where-why-how} framework, we discuss the implications of the use of word-scale graphics practices for scholarly communication. In the following, we first translate our findings in designing word-scale graphics for scholarly communication (\S\ref{sec:navigating}), and then reflect on the limitations of our study (\S\ref{sec:limitations}).

\subsection{Word-scale Graphics for Scholarly Communication}
\label{sec:navigating}

\subsubsection{Expanding the definition of word-scale graphic ``entities''}
We observed that word-scale graphics appear in both text and tables, serving distinct scholarly communication purposes, and have diverse visual representations. We argue that our findings further expand on the definition of word-scale graphic ``entities''. Previous works have defined an entity ``a concrete piece of text that has associated metadata'' serving as the ``link'' between graphic and text~\cite{goffin2014Exploring, goffin2017Exploratory}. In our corpus, we found that word-scale graphics are frequently used for visual indexing across text, tables, and figures. In these cases, the indexed target is not always textual: authors sometimes point to visual elements such as table structures or figure components. We also observe typography-based cases in which the text's styling provides the indexing cue, making the text simultaneously the entity and the word-scale graphic. Hence, we argue for expanding the notion of an ``entity'' beyond text--entities should encompass \textit{any expressive component within a document}. We advocate that researchers and practitioners design around this broader definition when considering the application of word-scale graphics.

\subsubsection{Promoting word-scale graphics use in scientific papers.}
Besides proposing an expanded design space for word-scale graphics by rethinking how we define entities, we also discuss the current barriers to their use in scientific papers, despite their advantages in fostering document understanding (\eg~\cite{beck2016Visual, goffin2015Exploring, zou2025GistVis}). Barriers exist in both the authoring and reading experiences with word-scale graphics, hindering their adoption. For instance, in our corpus, only 15.8\% of the identified cases are quantitative graphs, while most are icons. One possible explanation of such scarcity might stem from the author's concern about its practical limitations, given its small size. However, reflecting on the corpus, we argue that word-scale graphics might be especially useful as ``information scents'' to key information scattered across the paper. For data-driven word-scale graphics, while visualization of large datasets or complicated relations (\eg complex network graphs) might not be feasible, they can be helpful in providing additional context when the data is curated to be single variate (\eg trend lines for single variate time series data, scatter plot/violin plot to show a distribution). Another alternative hypothesis for this phenomenon is that creating data-driven word-scale graphics is difficult, leading to a preference for integrating easier word-scale graphic types. Future work could explore tools that streamline the creation process by linking word-scale graphic entities and representations, such as supporting dynamic text-data linkages \cite{zhu-tian2022CrossData} with generative LLM capabilities \cite{sobrien2024Inline}. On the side of reading, the current paper remains static even when presented via interactive formats (\ie HTML renders), and we join the effort in augmenting scholarly communication by further considering the interactivity of word-scale graphics in scientific papers through various toolkits (\eg~\cite{heer2023livingpapers}). 
Additionally, we also advocate for administrative innovations to improve infrastructure to enable authors to be more expressive through word-scale graphics. Current production workflows (\eg ACM TAPS\footnote{\scriptsize \url{https://authors.acm.org/proceedings/production-information/taps-production-workflow}}) impose strict restrictions on allowable LaTeX packages and dependencies to support cross-compilation to both PDF and HTML formats and support accessibility. We therefore advocate that researchers join forces to advocate for more flexible publication processes to rethink academic publishing and resolve challenges along the process.

\subsection{Limitations and Future Work}
\label{sec:limitations}
We outline the limitations of our study and identify directions for future work. First, our analysis is based on computer science papers from 2024. While this provides insight into current word-scale graphics use in computer science papers, it is currently unclear how these findings generalize beyond computer science or to other time periods. Second, our extraction is based on heuristics and may miss some use cases. Third, because our primary objective was to establish a framework for the use of word-scale graphics, we did not investigate how the word-scale graphics were created (\eg arXiv provides raw LaTeX source code). Future work could expand the corpus to include diverse fields and time periods, thereby refining our framework and further understanding the challenges of creating word-scale graphics in scientific papers by analyzing the LaTeX source code. Fourth, the categorization has room for future improvements. For instance, the serial mark type could also be categorized as a type of typography. While we eventually decided to categorize it under icons, future work with more evidence grounded in visualization theory could revise this categorization to make the framework more consistent. Moreover, cases might exist in which a word-scale graphic falls under multiple usage patterns. Lastly, while we present examples to the \textit{how} dimension, we did not discuss how other visualization taxonomies might be applied to icons or quantitative graphs. Future work should drill down further to characterize the visual representation of word-scale graphics and perform a more detailed analysis of their usage.

\section{Conclusion}
\label{sec:conclusion}

In this work, we developed a \textit{where-why-how} framework to characterize word-scale graphics use in the wild based on a corpus study of $126,797$ papers. We found that only a few papers employed word-scale graphics in scholarly communication. Moreover, papers using word-scale graphics are dominated by icons. By systematically proposing a framework for word-scale graphics, we aim to provide a ``language'' for researchers to characterize their use and, in turn, spark broader discussion on adopting word-scale graphics for scientific communication.

\section*{Acknowledgment of the Use of AI}
We acknowledge the use of Generative AI to improve the grammar, style, and readability of this manuscript. These tools were used solely for text editing, and played no role in the interpretation or generation of the core findings presented.

\begin{acks}
This work was supported by the National Innovation Training Program for College Students (Grant No.202510247067). We extend our thanks to the anonymous reviewers for their constructive feedback.
\end{acks}

\balance
\bibliographystyle{ACM-Reference-Format}
\bibliography{bib/website_ref,bib/zotero_export}

\end{document}